\documentclass{vgtc}                          

\ifpdf
  \pdfoutput=1\relax                   
  \usepackage{graphicx}                
  \DeclareGraphicsExtensions{.pdf,.png,.jpg,.jpeg} 
\else
  \usepackage{graphicx}                
  \DeclareGraphicsExtensions{.eps}     
\fi%

\graphicspath{{figures/}{pictures/}{images/}{./}} 

\usepackage{microtype}                 
\PassOptionsToPackage{warn}{textcomp}  
\usepackage{textcomp}                  
\usepackage{mathptmx}                  
\usepackage{times}                     
\usepackage{cite}                      
\usepackage{tabu}                      
\usepackage{booktabs}                  

\usepackage{subfig}

\onlineid{0}

\vgtccategory{Research}

\vgtcinsertpkg

\title{Why Shouldn’t All Charts Be Scatter Plots? \\ 
Beyond Precision-Driven Visualizations}

\author{Enrico Bertini\\ %
    \scriptsize New York University %
\and Michael Correll\\ %
    \scriptsize Tableau Research %
\and Steven Franconeri\\ %
    \scriptsize Northwestern University
    }

\teaser{
  \centering
  \subfloat[]{
   \includegraphics[width=0.25\textwidth]{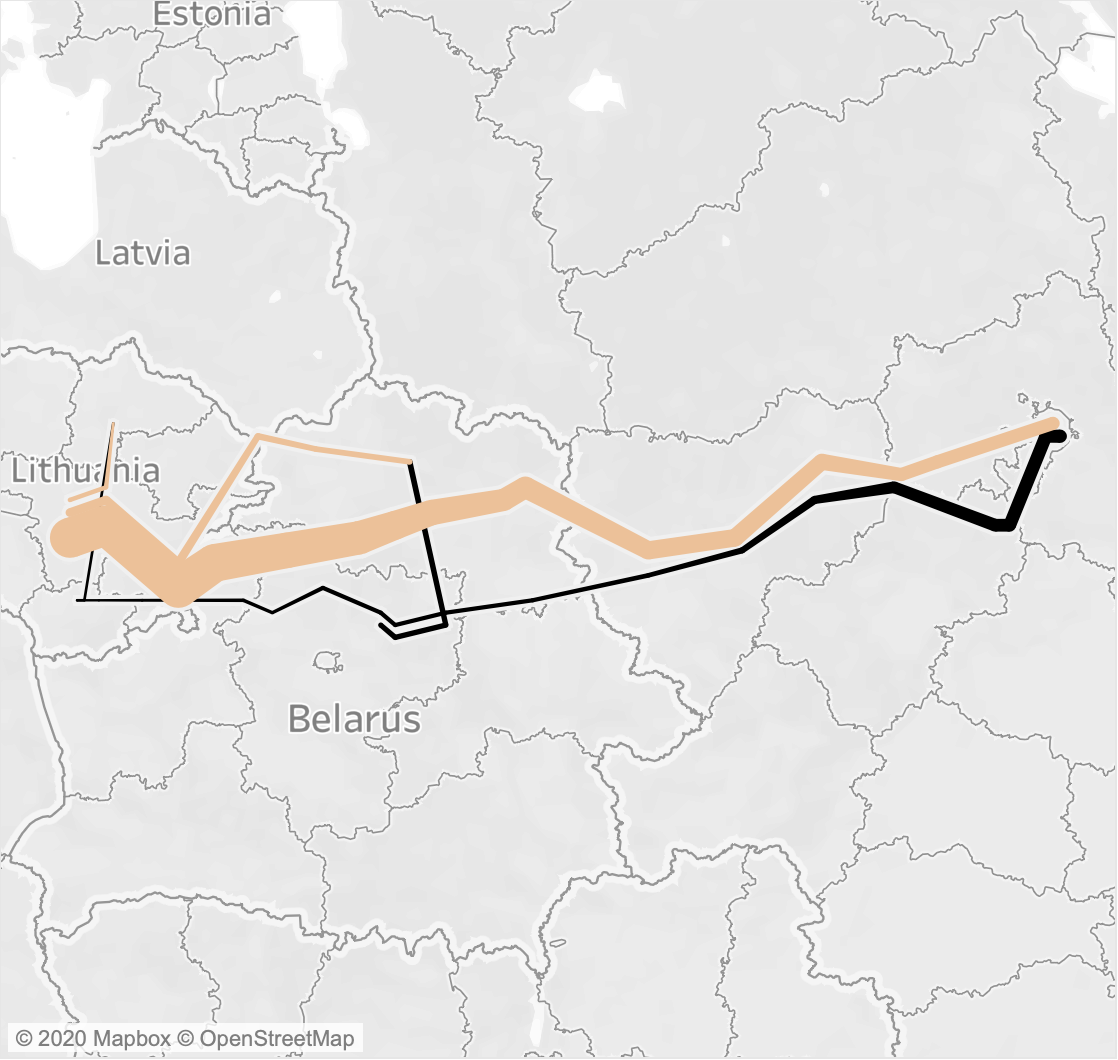}
   \label{fig:minard}
 }
 \subfloat[]{
   \includegraphics[width=0.45\textwidth]{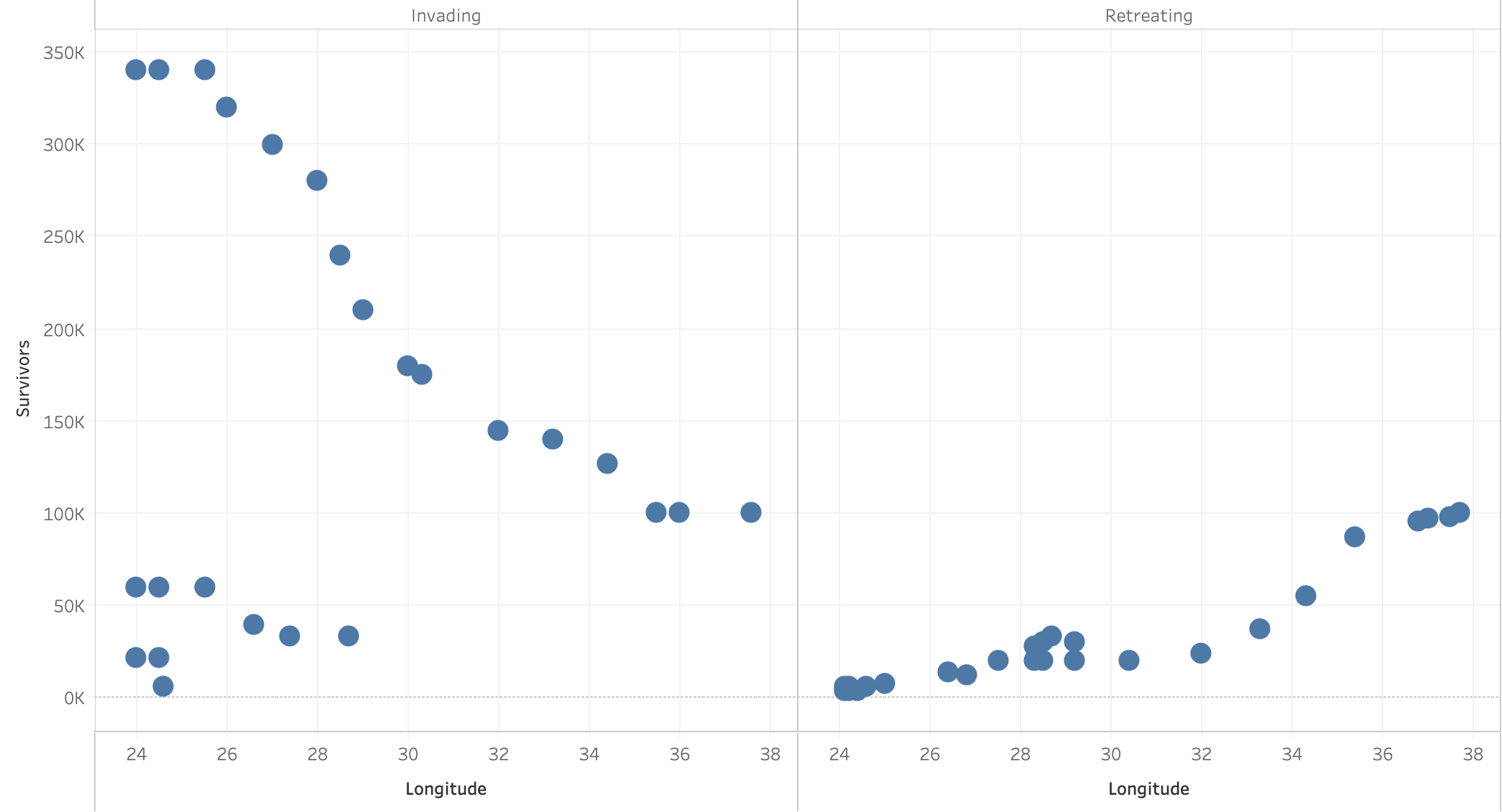}
   \label{fig:dotminard}
  }
  \caption{Two re-creations of Charles Minard's map of the invasion of Russia in Tableau. In (\ref{fig:minard}) we hew as closely as possible to Minard's design. In (\ref{fig:dotminard}) we attempt to encode all of the same data about the three groups of the invasion force using the most efficient channel: position on a common axis. In which circumstances might we prefer (\ref{fig:minard}) over (\ref{fig:dotminard})?}
}

\abstract{
A central concept in information visualization research and practice is the notion of visual variable \emph{effectiveness}, or the perceptual precision at which values are decoded given visual channels of encoding. Formative work from Cleveland \& McGill has shown  that position along a common axis is the most effective visual variable for comparing individual values. One natural conclusion is that \emph{any} chart that is not a dot plot or scatterplot is deficient and should be avoided. In this paper we refute a caricature of this ``scatterplots only'' argument as a way to call for new perspectives on how information visualization is researched, taught, and evaluated.
} 

\CCScatlist{
  \CCScatTwelve{Human-centered computing}{Visu\-al\-iza\-tion}{Visualization theory, concepts and paradigms}{}
}

\begin{document}



\firstsection{Introduction}

\maketitle

Minard's famous map of the Grande Arm\'ee's invasion of Russia has been called one of the ``best statistical drawings ever created''~\cite{tufte2001visual} and presents complex geographic, logistical, and weather data simultaneously. It is possible to recreate many aspects of Minard's map in common visualization systems like Tableau (\ref{fig:minard}). However, many of the encodings used in the map (such as width of lines with arbitrary, non-aligned angles) are comparatively imprecise for the estimation of values. A na\"ive recreation of the same data, based purely on the ``efficiency'' of visual variables, might look more like (\ref{fig:dotminard}). In (\ref{fig:dotminard}), the user can more precisely determine the size of individual groups of the Arm\'ee at specific time points. A quantitative evaluation of (\ref{fig:dotminard})'s performance (in terms of response time and accuracy) might very well find it to be superior to (\ref{fig:minard}) on that basis. And yet, the conclusion that (\ref{fig:dotminard}) is a strictly superior visualization, or that it would be equally iconic and compelling as Minard's map, seems unfounded. When might we prefer one version over the other, and what empirical evidence exists in the visualization literature to ground these preferences?

This example highlights an apparent contradiction at the heart of information visualization. On one hand, our exemplars of good visualizations can be diverse, complex, and reward contemplation\cite{heer2010tour}. On the other hand, our foundational empirical results and rules of thumb are often simple and minimalist. These rules are typically evaluated in terms of how quickly and accurately people extract specific information from charts, including formative psychophysical studies showing that viewers extract data values most precisely when they are encoded via position on shared axes ~\cite{cleveland1984graphical}.


Given these constraints, a natural conclusion is that \textit{quantitative data should almost always be depicted in a dot plot or scatter plot}, perhaps breaking data into SPLOMs, small multiples, or employing brushing and linking when there are too many variables for one view. While this argument is a strawman, its premises lie at the heart of foundational visualizations books by authors like Bertin and Tufte, and embodied as charts of rankings of visualization effectiveness in text books that are at the heart of how we teach visualization to students~\cite{munzner2014visualization,ware2019information}. In our own teaching we have struggled with how to convey these perceptual and design principles without resorting to at least some form of this argument. Therefore, we find this strawman useful to knock down: the goal of this paper is to highlight additional set of constraints that compete with perceptual precision, both in the mind of the designer and the studies of the researcher. We argue for a more expansive view of visualization beyond the perceptually precise encoding and decoding of individual data values, and make the case for ``inefficient'' visualizations. 



Our refutation of this argument focuses on attacking three hidden premises, critique of which reveals three classes of insufficiencies. The first premise is that accuracy in the extraction of data values is a sufficient measure of \textit{perceptual precision}. The second premise is that perceptual precision is a sufficient measure of a chart’s utility in \textit{communicating data}. The third premise is that utility in communicating data is sufficient to understand the larger \textit{purpose and power} of visualizations. We address these three premises in turn, using the insufficiencies associated with each to motivate a more expansive view of visualization design and pedagogy, and to suggest directions for future research.


\section{Beyond Individual Values}

\begin{table*}[]
\begin{tabular}{llc}
\textbf{Task} & \textbf{Example} & \textbf{Existing Work} \\
\hline
\textbf{1 pair of 2 points}           & \textit{Between months 1 and 2 in 2014...  }                                & \\
Metric: Retrieve value / two-point ratio & ...what is their ratio? & \cite{cleveland1984graphical,heer2010crowdsourcing,talbot2014four} \\
Ordinal: What's the relation of this pair? & ...which is bigger? & \\
\textbf{N pairs of 2 points}           & \textit{For months 1 and 2 across all years...  }                                & \\
Metric: N pairs largest ratio & ...which year had the largest ratio between values? & \cite{srinivasan2018s} \\
Metric: N pairs average ratio & ...what is the average ratio between months across all years? & \cite{nothelfer2019measures}\\
Metric: N pairs cluster into M relation types & ...what types of ordinal relations are present? &  \\
Ordinal: Are all ordinal relations present? & ...do they all fall, or do any rise? &  \\
Ordinal: Modal pair relation type  & ...is one ordinal relationship type most frequent? & \cite{nothelfer2019measures}\\
\textbf{N-point (patterns)}           & \textit{Across all months and years...  }                                & \\
Single point identification (min/max, outlier)  & ...which single data point is largest? & \cite{szafir2016four}\\
Summarization (mean, range)  & ...was 2011 or 2013 higher on average? & \cite{szafir2016four}\\
Cluster  & ...where would you place a threshold for low vs. high values? & \\
Extrapolation, imputation  & ...what is your guess for the value of the first month of 2016? & \cite{song2018s}\\
Shape, trend & ...how does the trend visually differ between 2012 and 2013? & \cite{correll2016semantics,gogolou2018comparing,lee2019you} \\
Filter & ...which year has the most near-zero values? & \\
Statistical tasks & ...which other year is most highly correlated with 2010? & \cite{rensink2010perception,harrison2014ranking,jardine2019perceptual,yang2018correlation}\\
\end{tabular}
\caption{A short list of potential value-related tasks in visualizations. Our recommendations for visual encodings and visualization designs are often related to only a narrow set of these tasks (often the first two, dealing with individual pairs of values). Empirical evidence about the precision of visual encodings for other tasks is often sparse or counter to existing recommendations.}
\label{tab:examples}
\end{table*}

\begin{figure*}[t!]
\centering
\includegraphics[width=0.9\textwidth]{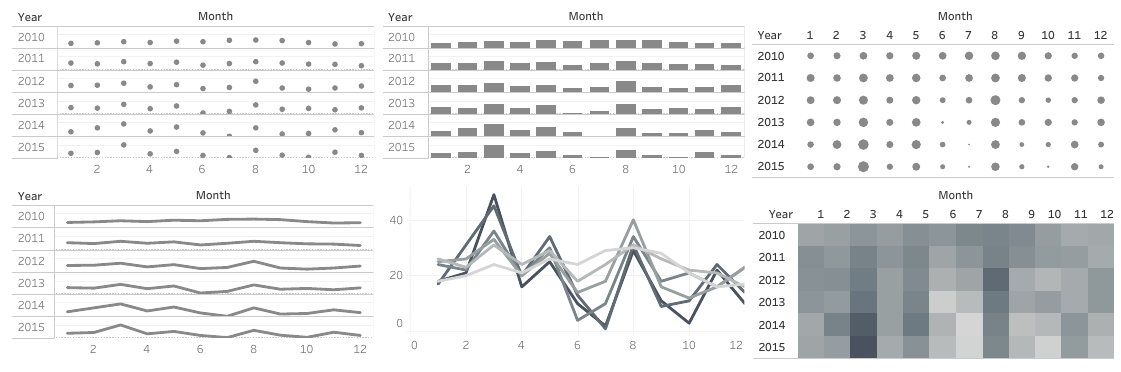}
\caption{Sample data for the tasks laid out in Table \protect\ref{tab:examples}. Consider which tasks are subjectively easier or harder across these different designs.
}
\label{fig:examplesFig}
\end{figure*}

Viewers are most efficient at computing the ratio between two visualized values when those values are encoded by their position on a common axis, as in a dot plot. The dominance of position encodings for this task is followed by a ranked list of other encodings including 2D area and orientation, and with intensity typically listed as the least precise encoding~\cite{cleveland1984graphical}. There is an implicit assumption that the ranking derived from this two-value ratio judgment represents an atomic unit for visualization, so that the additional precision conveyed by position should transfer to better perceptual performance in more complex tasks.

We challenge this assumption. While the precision of ratio judgments is one operationalization of perceptual performance, there are many others. For example, one fundamental analytics task is seeing the 'big picture' in a dataset. Figure~\ref{fig:examplesFig},  provides an example, showing a $6\times12$ grid of data values plotted as a dot plot (position), bar graph (position + length),  bubble chart (area), line graphs (juxtaposed and superposed), and heatmap. If precision of value extraction is all that matters, then the dot plot should be the preferred design. In contrast, it is clear to our eyes that the position-encoded dot plot is the \textit{least} effective visualization for seeing many potential 'big pictures' of the data. The bar graph to its right is far more useful for this task, likely because it adds a redundant encoding of length (or more likely, area \cite{yuan2019perceptual}), The line graph below is useful because it adds an emergent encoding of the local deltas between points via the orientation of the lines. Our favorite 'big picture view' is actually the heatmap in the corner, despite its status as the bottom of the barrel for precision of extracting ratios between individual values.

Table \ref{tab:examples} depicts a list of other likely perceptual tasks, inspired by work on low-level task taxonomies~\cite{amar2004best} \& recent papers that examine visualization through the lens of perceptual psychology~\cite{albers2014task,correll2012comparing,jardine2019perceptual,yang2018correlation}. We do not claim that this list is exhaustive, representative, or even correct. It is instead intended to show that ratio judgment tasks are only a small subset of likely perceptual tasks. The second column of Table \ref{tab:examples} provides concrete examples of the abstract perceptual tasks within that concrete example.

In the first two rows of the table, we list two perceptual operations that can be computed over 2 points: ‘metric’ relations between metric numbers (a ratio) and ‘ordinal’ relations between pairs of metric values (is value A higher than B?), say the first two points of the first row of Figure~\ref{fig:examplesFig}. While data visualization research has focused almost exclusively on the former, the latter is arguably as important for real-world tasks. While we occasionally note that today is five degrees hotter than yesterday, we more typically note that it is hotter than yesterday. COVID-19 infection rates have increased. Profits are lower, and we are over budget. 

The next set of rows list tasks that might unfold when a viewer is presented with many 2-point pairs of values, such as the first two rows of one of the visualizations in Figure~\ref{fig:examplesFig}. These tasks include metric comparisons, such as finding which pair has the largest ratio, estimating an average ratio, or clustering the sizes of ratios. They also include ordinal comparisons of metric value pairs, such as finding a unique relation ($A\leq B$ among $A\geq B$ pairs), or estimating which relations type is more frequent. We know of only two studies that have studied these important perceptual tasks~\cite{nothelfer2019measures,srinivasan2018s}, but because both rely primarily on position encodings, they cannot confirm whether the Cleveland \& McGill position ranking holds for these alternative tasks. 

The next several rows show perceptual tasks that are not constrained over pairs of points, and instead could be computed over an entire set or subset of N values. These include identifying a single value with a given property (e.g., min, max, outlier), or summarizing a set of values by a single number (e.g., mean, variance, clusters). Recent work on the perception of these ``aggregate'' or ``ensemble'' tasks~\cite{albers2014task,szafir2016four} provides evidence that many encodings that are imprecise for individual values (such as color) have performance benefits for these types of tasks over positional encodings like line charts.

The row labeled 'Shape, trend' refers to the need to holistically judge a single series, or to compare two series, in an open-ended manner. We suspect that this task, like stepping back to see the 'big picture' in the data, will not always be best supported by position encodings. The viewer might search for anything from basic patterns (rising, flat), to idiosyncratic motifs and shapes in the data~\cite{gogolou2018comparing}. Visual interfaces for times series search~\cite{lee2019you} have had to find ways for users to express shapes (and the properties of those shapes that they find important~\cite{correll2016semantics}) in fluid and dynamic ways, as the rigid definition of specific individual values may not capture the visual features of interest to the user.

The row labeled 'filter' refers to a visual subset operation based on the data values, e.g., ‘pick out all of the high values’. We do not have a full understanding of the filtering operation in visualization contexts, although existing work evaluates the detection of individual ``oddball'' outliers~\cite{haroz2012capacity} or filtering across nominal categories (for instance, picking a particular class of points out of a scatterplot~\cite{gleicher2013perception}). However, perceptually motivated designs for time series data have often used color as a form of perceptual ``boosting''~\cite{oelke2011visual} to highlight anomalous items~\cite{albers2011sequence,correll2015layercake}.

Finally, recent work has begun to uncover the perceptual tasks that underlie more complex comparisons, such as judging the correlation between two sets of paired values~\cite{rensink2010perception,harrison2014ranking,yang2018correlation}. This work suggests that, instead of judging high-level properties like correlation \textit{per se}, a viewer relies on a more concrete proxy, such as the aspect ratio of the bounding box surrounding the points~\cite{yang2018correlation}. These hypothesized proxies may prove to be complex and may take many years to unpack  -- but when they are better understood, we cannot predict whether they would be best supported by position encodings. 

\section{Metaphors and Congruence}
The choice of encoding channels is constrained by more than perceptual precision, with one major constraint being the congruency of its metaphors~\cite{mackinlay1986automating}. A channel must be consistent with (and convey) the concepts that they encode -- for example, conveying quantity with area -- serving as type of ``affordance" for the data to show how it can be used (e.g., a push plate vs. a pull-bar for a door) ~\cite{norman2014things}.




One example of this conceptual congruence is a study that asked participants to describe simple bar and line graphs of the same two data points ~\cite{zacks1999bars}. Bar charts lead to descriptions in terms of discrete comparisons whereas lines lead to descriptions of trends; indicating that different graphical solutions can suggest different type of interpretations. Interestingly, in these examples the channels used to convey quantitative information were identical (i.e., vertical position) and the only element that changes between the graphs is the affordance of the connected line implying continuous data, and the visual connection between points in the line graph. 


Another example comes from one author's experience with an exercise assigned in his information visualization class. The assignment asked the students to compare a set of countries in terms of amount of money donated or received, as recorded in the Aid Data data set \footnote{https://www.aiddata.org/} (which records international aids disbursements globally). Students produced two main type of solutions: (1) a scatter plot with dots representing the countries and axes representing incoming and outgoing amounts; (2) a pair of aligned bar charts showing for each country the amount donated and the amount received. While both solutions employ position as the main channel to encode the donated/received amounts, the graphs invite the reader in making completely different sets of judgments. More precisely, while the scatter plot affords detection of correlations and groupings, the pair of aligned bar charts invites the reader to compare individual countries across two metrics (see Fig. ~\ref{fig:examplesFig} for a similar design comparison). These examples show that the ranking of visual channels (based on precision) is not \textit{sufficient} in the visualization design space. In other words, knowing that a channel affords more precision does not provide sufficient guidance for visualization design.


The role of metaphors and expressiveness can be observed at multiple levels of granularity. At the level of individual channels there are several examples of how channels may express or fail to express certain types of information. For example, color hue can’t express ordinal or quantitative information because the human eye does not assign an order to colors that vary exclusively in hue. Similarly, colors have strong semantic associations, therefore appropriate associations between concepts and colors may improve readability and comprehension~\cite{lin2013selecting}. Area size is a moderately precise channel when conveying quantity, but it cannot easily show negative values because larger sizes are firmly associated with larger (positive) quantities. 



Different semantic associations can also be created by using different symbols or graphical marks. A classic example is the representation of part-to-whole relationships and the question of whether pie charts should be considered effective solutions for the representation of such data~\cite{skau2016arcs}. When in a visualization the designer wants to explicitly convey information about the fact that a given value is part of a whole, specific metaphors work better than others. For example, in comparing a pie chart, a stacked bar and a group of bars, it is evident that only the pie chart and the stacked bar explicitly convey the part-to-whole metaphor. Following the reasoning behind the ranking of visual variables, the solution with separate bars (position encoding) should be preferred over the stacked bar (length encoding) or pie chart (angle and area encoding) because it provides a more precise representation. Other similar examples of this kind exist. For instance, bars on maps are rarely used, whereas circles are often preferred in their place. Line charts are preferred over bars when the goal is to convey a temporal trend. Icon arrays are preferred over aggregate values in risk estimation. All of these examples demonstrate that there is something more than ranking of visual channels and that reasoning about visualization at the level of individual channels can be limited and potentially misleading.





Even a combination of accuracy and efficiency cannot fully characterize the effectiveness of data visualizations. One must also measure how easy it is to extract information out of it.

Two concepts developed in the literature on cognitive science seem to be of pertinence here. The first one is the ``congruence principle'' suggested by Tversky et al.~\cite{tversky2002animation}. The principle states that ``\textit{the content and format of the graphic should correspond to the content and format of the concepts to be conveyed}'' and it seems to apply perfectly to the type of concerns we discussed above. The second concept is ``cognitive fit,'' developed by Vessey. In the words of Vessey~\cite{vessey1991cognitive}: ``\textit{... performance on a task will be enhanced when there is a cognitive fit (match) between the information emphasized in the representation type and that required by the task type.}'' While the theory of cognitive fit has been developed originally to explain the difference between symbolic and graphical representations (i.e., tables vs graphs), there is no reason to believe the same logic can’t be used to describe differences between alternate graphical representations. A good matching between the ``information emphasized in the representation type" and the information a reader is expected to extract seems to be a good guiding principle for data visualization.

\section{Rhetoric, Persuasion, and Memory}
A last category of objections to a world of only scatterplots is that many visualizations are unconcerned with accurate extraction of individual (or even aggregate) values. Charts are often designed to persuade, educate, and motivate. Designing for serendipitous discovery, educational impact, hedonic response, or changes in behavior is in some cases only tangentially connected with the precision of a particular visualization. Wang et al.~\cite{wang2019emotional} call for us to ``revis[e] the way we value visualizations''  on this basis, and Correll \& Gleicher~\cite{correll2014bad} point to a whole class of designs and design guidelines that seem counter-productive in terms of precision but that nonetheless result in benefits in terms of higher-level cognitive goals. In this section we briefly discuss some of these mismatches.

Hullman et al.~\cite{hullman2011benefitting} point to possible benefits for ``visual difficulties’’ in charts: that is, by making the viewer do more work to decode the values, there is potentially an impact on the retention of those values. Rather than designing charts to be as precise as possible, for longer-term or higher-level tasks we may wish to slow the viewer down. Lupi’s~\cite{lupi2017data} Data Humanism manifesto calls for visualizations that encourage viewers to ``spend time’’ with the data, with examples of dense, multidimensional glyph-based visualizations that do not afford quick and precise extraction of values. Similarly, Bradley et al.~\cite{bradley2019approaching} call for a ``slow analytics’’ movement that encourages ownership and retention of analytical tasks rather than precision.

Part of the pedagogical utility of charts is not merely conveying the information, but ensuring that the information is retained. One immediate downside to a world of only scatter and dot plots is that our charts would all look similar, and so unlikely to be differentiated much in memory. Borkin et al.~\cite{borkin2013makes,borkin2015beyond} find that charts with pictorial elements and other visual features of interest are more memorable than plain and otherwise unadorned charts. Kostelnick~\cite{kostelnick2008visual} recommends occasional deviations from minimalist design in the service of ``clarity'' which can include such factors as engaging the reader's attention. Many of the most impactful charts in visualization have had non-standard or otherwise less than precise forms (e.g., Fig~\ref{fig:minard}).

There may be benefits for imprecise visualizations for analysts as well, not just for passive viewers or learners. Often when designing a system we may have no idea of the form or category of insights present in our data. The serendipitous discovery of important features of the data-set may not be well-covered by existing design principles that are designed for the precision at intended or standard analytical tasks: lucky, chance, and stochastic exploration may be more important than reliably picking out values. Thudt et al.~\cite{thudt2012bohemian} discuss the challenges of designing for serendipitous information discovery, and suggest that standard designs may be ill-suited to the unconstrained and stochastic sort of exploration that can be necessary for making discoveries, whereas D{\"o}rk et al.~\cite{dork2011information} point to the challenges of designing for the wandering ``information flâneur.''

There are also potentially \textit{costs} to overly-precise visualizations. Kennedy et al.~\cite{kennedy2016work} claim that the ``clean layouts'' of minimalist visualizations can grant an imprimatur of authority and objectivity to data that may not match that standard. Likewise, Drucker~\cite{drucker2012humanistic} points to the ``seductive rhetorical force'' of visualizations to convince viewers that the data they contain is not merely a potentially flawed, biased, and uncertain view, but an objective truth about the world. This unwillingness to question charts due to a perception of their objectivity can override even strong political conventions or skepticism~\cite{peck2019data}. ``Messier'' designs (such as sketchy~\cite{wood2012sketchy} or uncertainty-conveying~\cite{song2018s} renderings) can introduce a willingness to critique or a greater appreciation for uncertainty not present in more precision-driven visualizations.    

\section{Conclusion}
Much of the empirical and theoretical basis for visualization work comes from studies examining the efficiency of visual channels in extracting information, and using these results to generate a ranking of these channels~\cite{cleveland1984graphical}. These rankings power many of our design guidelines and constraints~\cite{moritz2018formalizing}, are ubiquitous in our textbooks~\cite{munzner2014visualization, ware2019information}, and are instantiated in the logic of many of our automated or semi-automated visualization design tools~\cite{mackinlay1986automating,mackinlay2007show}. And yet, these rankings do not seem to capture important components of how people use, interpret, and learn from visualizations. We should be expansive in how we analyze, conceptualize, and teach visualization. Otherwise, we risk a situation where academia focuses on the narrow, scatterplot-like section of the vast, more interesting world of visualization as a whole.

Of course we are not proposing to throw the baby out with the bathwater. The ranking of visual variables has had enormous impact on visualization research and practice, informing design decisions for tool development and providing pedagogical value in numerous guidelines, textbooks, and courses. Our intent is to raise awareness about the ways an excessive and narrow focus on channel ranking may be acting as a detrimental limitation to our field in terms of: (1) understanding actual data visualization practice; (2) development of data visualization tools and techniques; (3) methodologies for data visualization design and evaluation and (4) pedagogy of data visualization. The question is: how can we rectify and expand the theory behind the ranking of visual variables? When does it work? When does it not work? And, maybe even more importantly, what else to do we need in its place or in addition to it? From this initial analysis of the various insufficiencies we have identified it seems clear there is much to do in this area. It is our hope that this work sparks interesting conversations and potentially lead other practitioners and designers to develop alternative (or more refined) practices, conceptualizations, and epistemologies~\cite{meyer2019criteria} for visualization.

\acknowledgments{
This work was supported by NSF awards IIS-1901485 and IIS-1900941.
}

\bibliographystyle{abbrv-doi}

\bibliography{references}

\begin{thebibliography}{10}

\bibitem{albers2014task}
D.~Albers, M.~Correll, and M.~Gleicher.
\newblock Task-driven evaluation of aggregation in time series visualization.
\newblock In {\em Proceedings of the SIGCHI Conference on Human Factors in
  Computing Systems}, pp. 551--560, 2014.

\bibitem{albers2011sequence}
D.~Albers, C.~Dewey, and M.~Gleicher.
\newblock Sequence surveyor: Leveraging overview for scalable genomic alignment
  visualization.
\newblock {\em IEEE transactions on visualization and computer graphics},
  17(12):2392--2401, 2011.

\bibitem{amar2004best}
R.~Amar and J.~Stasko.
\newblock A knowledge task-based framework for design and evaluation of
  information visualizations.
\newblock In {\em IEEE Symposium on Information Visualization}, pp. 143--150.
  IEEE, 2004.

\bibitem{borkin2015beyond}
M.~A. Borkin, Z.~Bylinskii, N.~W. Kim, C.~M. Bainbridge, C.~S. Yeh, D.~Borkin,
  H.~Pfister, and A.~Oliva.
\newblock Beyond memorability: Visualization recognition and recall.
\newblock {\em IEEE transactions on visualization and computer graphics},
  22(1):519--528, 2015.

\bibitem{borkin2013makes}
M.~A. Borkin, A.~A. Vo, Z.~Bylinskii, P.~Isola, S.~Sunkavalli, A.~Oliva, and
  H.~Pfister.
\newblock What makes a visualization memorable?
\newblock {\em IEEE Transactions on Visualization and Computer Graphics},
  19(12):2306--2315, 2013.

\bibitem{bradley2019approaching}
A.~J. Bradley, V.~Sawal, and C.~Collins.
\newblock Approaching humanities questions using slow visual search interfaces.
\newblock In {\em IEEE 4th Workshop for Visualization and the Digital
  Humanities}, 2019.

\bibitem{cleveland1984graphical}
W.~S. Cleveland and R.~McGill.
\newblock Graphical perception: Theory, experimentation, and application to the
  development of graphical methods.
\newblock {\em Journal of the American statistical association},
  79(387):531--554, 1984.

\bibitem{correll2012comparing}
M.~Correll, D.~Albers, S.~Franconeri, and M.~Gleicher.
\newblock Comparing averages in time series data.
\newblock In {\em Proceedings of the SIGCHI Conference on Human Factors in
  Computing Systems}, pp. 1095--1104, 2012.

\bibitem{correll2015layercake}
M.~Correll, A.~L. Bailey, A.~Sarikaya, D.~H. O’Connor, and M.~Gleicher.
\newblock Layercake: a tool for the visual comparison of viral deep sequencing
  data.
\newblock {\em Bioinformatics}, 31(21):3522--3528, 2015.

\bibitem{correll2014bad}
M.~Correll and M.~Gleicher.
\newblock Bad for data, good for the brain: Knowledge-first axioms for
  visualization design.
\newblock In {\em DECISIVe : Workshop on Dealing with Cognitive Biases in
  Visualisations}, 2014.

\bibitem{correll2016semantics}
M.~Correll and M.~Gleicher.
\newblock The semantics of sketch: Flexibility in visual query systems for time
  series data.
\newblock In {\em 2016 IEEE Conference on Visual Analytics Science and
  Technology (VAST)}, pp. 131--140. IEEE, 2016.

\bibitem{dork2011information}
M.~D{\"o}rk, S.~Carpendale, and C.~Williamson.
\newblock The information flaneur: A fresh look at information seeking.
\newblock In {\em Proceedings of the SIGCHI conference on human factors in
  computing systems}, pp. 1215--1224, 2011.

\bibitem{drucker2012humanistic}
J.~Drucker.
\newblock Humanistic theory and digital scholarship.
\newblock {\em Debates in the digital humanities}, pp. 85--95, 2012.

\bibitem{gleicher2013perception}
M.~Gleicher, M.~Correll, C.~Nothelfer, and S.~Franconeri.
\newblock Perception of average value in multiclass scatterplots.
\newblock {\em IEEE transactions on visualization and computer graphics},
  19(12):2316--2325, 2013.

\bibitem{gogolou2018comparing}
A.~Gogolou, T.~Tsandilas, T.~Palpanas, and A.~Bezerianos.
\newblock Comparing similarity perception in time series visualizations.
\newblock {\em IEEE transactions on visualization and computer graphics},
  25(1):523--533, 2018.

\bibitem{haroz2012capacity}
S.~Haroz and D.~Whitney.
\newblock How capacity limits of attention influence information visualization
  effectiveness.
\newblock {\em IEEE Transactions on Visualization and Computer Graphics},
  18(12):2402--2410, 2012.

\bibitem{harrison2014ranking}
L.~Harrison, F.~Yang, S.~Franconeri, and R.~Chang.
\newblock Ranking visualizations of correlation using weber's law.
\newblock {\em IEEE transactions on visualization and computer graphics},
  20(12):1943--1952, 2014.

\bibitem{heer2010crowdsourcing}
J.~Heer and M.~Bostock.
\newblock Crowdsourcing graphical perception: using mechanical turk to assess
  visualization design.
\newblock In {\em Proceedings of the SIGCHI conference on human factors in
  computing systems}, pp. 203--212, 2010.

\bibitem{heer2010tour}
J.~Heer, M.~Bostock, and V.~Ogievetsky.
\newblock A tour through the visualization zoo.
\newblock {\em Communications of the ACM}, 53(6):59--67, 2010.

\bibitem{hullman2011benefitting}
J.~Hullman, E.~Adar, and P.~Shah.
\newblock Benefitting infovis with visual difficulties.
\newblock {\em IEEE Transactions on Visualization and Computer Graphics},
  17(12):2213--2222, 2011.

\bibitem{jardine2019perceptual}
N.~Jardine, B.~D. Ondov, N.~Elmqvist, and S.~Franconeri.
\newblock The perceptual proxies of visual comparison.
\newblock {\em IEEE Transactions on Visualization and Computer Graphics},
  26(1):1012--1021, 2019.

\bibitem{kennedy2016work}
H.~Kennedy, R.~L. Hill, G.~Aiello, and W.~Allen.
\newblock The work that visualisation conventions do.
\newblock {\em Information, Communication \& Society}, 19(6):715--735, 2016.

\bibitem{kostelnick2008visual}
C.~Kostelnick.
\newblock The visual rhetoric of data displays: The conundrum of clarity.
\newblock {\em IEEE Transactions on Professional Communication},
  51(1):116--130, 2008.

\bibitem{lee2019you}
D.~J.-L. Lee, J.~Lee, T.~Siddiqui, J.~Kim, K.~Karahalios, and A.~Parameswaran.
\newblock You can't always sketch what you want: Understanding sensemaking in
  visual query systems.
\newblock {\em IEEE transactions on visualization and computer graphics},
  26(1):1267--1277, 2019.

\bibitem{lin2013selecting}
S.~Lin, J.~Fortuna, C.~Kulkarni, M.~Stone, and J.~Heer.
\newblock Selecting semantically-resonant colors for data visualization.
\newblock {\em Computer Graphics Forum}, 32(3pt4):401--410, 2013.

\bibitem{lupi2017data}
G.~Lupi.
\newblock Data humanism: the revolutionary future of data visualization.
\newblock {\em Print Magazine}, 30, 2017.

\bibitem{mackinlay1986automating}
J.~Mackinlay.
\newblock Automating the design of graphical presentations of relational
  information.
\newblock {\em Acm Transactions On Graphics (Tog)}, 5(2):110--141, 1986.

\bibitem{mackinlay2007show}
J.~Mackinlay, P.~Hanrahan, and C.~Stolte.
\newblock Show me: Automatic presentation for visual analysis.
\newblock {\em IEEE transactions on visualization and computer graphics},
  13(6):1137--1144, 2007.

\bibitem{meyer2019criteria}
M.~Meyer and J.~Dykes.
\newblock Criteria for rigor in visualization design study.
\newblock {\em IEEE transactions on visualization and computer graphics},
  26(1):87--97, 2019.

\bibitem{moritz2018formalizing}
D.~Moritz, C.~Wang, G.~L. Nelson, H.~Lin, A.~M. Smith, B.~Howe, and J.~Heer.
\newblock Formalizing visualization design knowledge as constraints: Actionable
  and extensible models in draco.
\newblock {\em IEEE transactions on visualization and computer graphics},
  25(1):438--448, 2018.

\bibitem{munzner2014visualization}
T.~Munzner.
\newblock {\em Visualization analysis and design}.
\newblock CRC press, 2014.

\bibitem{norman2014things}
D.~Norman.
\newblock {\em Things that make us smart: Defending human attributes in the age
  of the machine}.
\newblock Diversion Books, 2014.

\bibitem{nothelfer2019measures}
C.~Nothelfer and S.~Franconeri.
\newblock Measures of the benefit of direct encoding of data deltas for data
  pair relation perception.
\newblock {\em IEEE transactions on visualization and computer graphics},
  26(1):311--320, 2019.

\bibitem{oelke2011visual}
D.~Oelke, H.~Janetzko, S.~Simon, K.~Neuhaus, and D.~A. Keim.
\newblock Visual boosting in pixel-based visualizations.
\newblock In {\em Computer Graphics Forum}, vol.~30, pp. 871--880. Wiley Online
  Library, 2011.

\bibitem{peck2019data}
E.~M. Peck, S.~E. Ayuso, and O.~El-Etr.
\newblock Data is personal: Attitudes and perceptions of data visualization in
  rural pennsylvania.
\newblock In {\em Proceedings of the 2019 CHI Conference on Human Factors in
  Computing Systems}, pp. 1--12, 2019.

\bibitem{rensink2010perception}
R.~A. Rensink and G.~Baldridge.
\newblock The perception of correlation in scatterplots.
\newblock {\em Computer Graphics Forum}, 29(3):1203--1210, 2010.

\bibitem{skau2016arcs}
D.~Skau and R.~Kosara.
\newblock Arcs, angles, or areas: Individual data encodings in pie and donut
  charts.
\newblock {\em Computer Graphics Forum}, 35(3):121--130, 2016.

\bibitem{song2018s}
H.~Song and D.~A. Szafir.
\newblock Where's my data? evaluating visualizations with missing data.
\newblock {\em IEEE transactions on visualization and computer graphics},
  25(1):914--924, 2018.

\bibitem{srinivasan2018s}
A.~Srinivasan, M.~Brehmer, B.~Lee, and S.~M. Drucker.
\newblock What's the difference? evaluating variations of multi-series bar
  charts for visual comparison tasks.
\newblock In {\em Proceedings of the 2018 CHI Conference on Human Factors in
  Computing Systems}, pp. 1--12, 2018.

\bibitem{szafir2016four}
D.~A. Szafir, S.~Haroz, M.~Gleicher, and S.~Franconeri.
\newblock Four types of ensemble coding in data visualizations.
\newblock {\em Journal of vision}, 16(5):11--11, 2016.

\bibitem{talbot2014four}
J.~Talbot, V.~Setlur, and A.~Anand.
\newblock Four experiments on the perception of bar charts.
\newblock {\em IEEE transactions on visualization and computer graphics},
  20(12):2152--2160, 2014.

\bibitem{thudt2012bohemian}
A.~Thudt, U.~Hinrichs, and S.~Carpendale.
\newblock The bohemian bookshelf: supporting serendipitous book discoveries
  through information visualization.
\newblock In {\em Proceedings of the SIGCHI Conference on Human Factors in
  Computing Systems}, pp. 1461--1470, 2012.

\bibitem{tufte2001visual}
E.~R. Tufte.
\newblock {\em The visual display of quantitative information}, vol.~2.
\newblock Graphics press Cheshire, CT, 2001.

\bibitem{tversky2002animation}
B.~Tversky, J.~B. Morrison, and M.~Betrancourt.
\newblock Animation: can it facilitate?
\newblock {\em International journal of human-computer studies},
  57(4):247--262, 2002.

\bibitem{vessey1991cognitive}
I.~Vessey.
\newblock Cognitive fit: A theory-based analysis of the graphs versus tables
  literature.
\newblock {\em Decision sciences}, 22(2):219--240, 1991.

\bibitem{wang2019emotional}
Y.~Wang, A.~Segal, R.~Klatzky, D.~F. Keefe, P.~Isenberg, J.~Hurtienne,
  E.~Hornecker, T.~Dwyer, and S.~Barrass.
\newblock An emotional response to the value of visualization.
\newblock {\em IEEE computer graphics and applications}, 39(5):8--17, 2019.

\bibitem{ware2019information}
C.~Ware.
\newblock {\em Information visualization: perception for design}.
\newblock Morgan Kaufmann, 2019.

\bibitem{wood2012sketchy}
J.~Wood, P.~Isenberg, T.~Isenberg, J.~Dykes, N.~Boukhelifa, and A.~Slingsby.
\newblock Sketchy rendering for information visualization.
\newblock {\em IEEE Transactions on Visualization and Computer Graphics},
  18(12):2749--2758, 2012.

\bibitem{yang2018correlation}
F.~Yang, L.~T. Harrison, R.~A. Rensink, S.~L. Franconeri, and R.~Chang.
\newblock Correlation judgment and visualization features: A comparative study.
\newblock {\em IEEE Transactions on Visualization and Computer Graphics},
  25(3):1474--1488, 2018.

\bibitem{yuan2019perceptual}
L.~Yuan, S.~Haroz, and S.~Franconeri.
\newblock Perceptual proxies for extracting averages in data visualizations.
\newblock {\em Psychonomic bulletin \& review}, 26(2):669--676, 2019.

\bibitem{zacks1999bars}
J.~Zacks and B.~Tversky.
\newblock Bars and lines: A study of graphic communication.
\newblock {\em Memory \& cognition}, 27(6):1073--1079, 1999.

\end{thebibliography}
\end{document}